\documentstyle[12pt]{article}
\begin{document}

\baselineskip=20pt

\null\vspace{2.0cm}
\begin{flushright}
TUIMP-TH-95/74\\
\end{flushright}

\vspace{0.5cm}
\begin{center}
{\Large\sf Heavy Fermion Screening Effects and Gauge Invariance}

\vspace{1.5cm}
{\bf Yi Liao$^b$ \hspace{1cm} Yu-Ping Kuang$^{a,b}$ \hspace{1cm} Chong-Sheng
Li$^{a,c}$}\\

\vspace{1.5ex}
\noindent
{\small {\it a China Center of Advanced Science and Technology (World
Laboratory),\\ PO Box 8730, Beijing 100080, China}\\
{\it b Institute of Modern Physics, Tsinghua University, Beijing 100084,
China}\footnote{Mailing address}\\
{\it c Department of Physics, Peking University, Beijing 100871, China}}\\
\vspace{5.0ex}

\normalsize
             {\bf Abstract}
\end{center}
We show that the naively expected large virtual heavy fermion effects
in low energy processes may be screened if the process under consideration
contains external gauge bosons constrained by gauge invariance. We illustrate
this by a typical example of the process $\gamma\gamma\to b \bar{b}$.
Phenomenological implications are also briefly indicated.
\begin{flushleft}
\end{flushleft}

\newpage
Studying the effects of heavy particles in radiative corrections is of special
importance for exploring new physics if the accelerator energy is not
sufficient to directly produce them. With respect to these effects, there
are two kinds of theories. In renormalizable theories with coupling constants
independent of the heavy particle masses like Quantum Electrodynamics (QED),
these effects are not significant since decoupling theorem$\cite{ac}$ shows
that the heavy particles completely decouple from the low energy physics
in the heavy mass limit. In non-decoupling theories to which the decoupling
theorem does not apply, these effects can be significant and are thus important
for studying new physics. A typical example of the non-decoupling theory is
the Standard Model of the electroweak theory (SM), in which heavy particles
may affect the low energy physics in two separate ways. Firstly, the heavy
top quark is a necessary ingredient in chiral anomaly cancellation, and
integrating it out will induce an effective Wess-Zumino-Witten term at low
energies$\cite{dhf}$, which is constant in heavy top limit. Secondly,
particles in the SM acquire mass from the fixed vacuum expectation value,
so that the heavy masses are proportional to the corresponding coupling
constants, and thus the conditions for the validity of the decoupling
theorem are not satisfied. This kind of non-decoupling can make certain
observables depend on positive powers of the heavy particle masses which
will blow up in the heavy mass limit. A well-known example is the one-loop
heavy top correction to the parameter $\rho$ reflecting the $W$, $Z$ boson
mass relation, which behaves as $G_F m_t^2$ $\cite{veltman}$ and originates
from the custodial $SU(2)$ symmetry $\cite{sikivie}$ breaking by the large
mass splitting between the top and bottom quarks. In the Higgs sector,
however, a similar correction from a heavy Higgs is absent due to Veltman's
screening theorem$\cite{screening}$. The naively expected leading terms of
$O(G_F m_H^2)$ at one-loop $\cite{screening}$ and $O(G_F^2 m_H^4)$ at two-
loop $\cite{bij}$ are cancelled in the $W$, $Z$ mass relation, and the
survivals are the next-to-leading terms of $O(G_F M_W^2 \ln m_H^2)$ and
$O(G_F^2 M_W^2 m_H^2)$, respectively. This phenomenon has been attributed in
Ref.$\cite{einhorn}$ to the vestige of the global custodial symmetry, and
generalized to all orders in perturbation theory.

In this paper, we shall point out that screening effects may also
appear in the heavy fermion sector if the low energy process under
consideration contains external gauge bosons which are constrained by gauge
invariance. As a result, the naively expected large correction to the process
from a virtual heavy fermion is actually small. Our discussion is based on a
simple analysis of gauge invariance and dimension counting. Although we take
the process $\gamma\gamma\to b \bar{b}$ as an example to illustrate the
screening effect of the large top mass, which is of interest by itself in
photon collider physics$\cite{brodsky}$, the whole analysis applies to the
general cases involving heavy fermions. We shall also briefly discuss the
processes $H\to \gamma\gamma$, $b\to s\gamma$ and indicate the
phenomenological implications.

At tree level, $\gamma\gamma\to b \bar{b}$ is a pure QED process.
In the following, we first focus on its one-loop correction arising from a
virtual heavy top and then generalize it to higher loops. As a theoretical
study, we are only interested in the leading-$m_t$ term corresponding to the
heavy top limit. Whether this is a good approximation is an issue
of phenomenology which is not the main purpose of this paper. In this limit we
may set the bottom mass to zero, $m_b=0$. We work in the $R_{\xi}$ gauge. The
leading term is contributed by the exchange of the unphysical Goldstone boson
$\phi^{\pm}$ ( and at higher loops by the exchange of the unphysical Goldstone
boson $\phi^0$ and physical Higgs boson $H$ as well ). The non-leading terms
which are of the same order as those from the ordinary electroweak
corrections are ignored here. Note that the non-leading terms are $\xi_{W(Z)}-$
dependent, and this dependence is cancelled only when corrections from $W$,
$Z$ bosons are included. With this consideration, the relevant interaction
Lagrangian at one-loop level is
\begin{equation}
\begin{array}{l}
{\cal L}_1=\displaystyle{\frac{g_2 m_t}{\sqrt{2}M_W} (\phi^+ {\bar t}_R b_L
                                        +\phi^- {\bar b}_L t_R)
           +eA_{\mu} (Q_t \bar{t}\gamma^{\mu}t
                     +Q_b \bar{b}\gamma^{\mu}b)} \\ \\

{}~~~~~\displaystyle{+ieA_{\mu}(\phi^- \partial^{\mu}\phi^+
                     -\phi^+ \partial^{\mu}\phi^-)
           +e^2 A_{\mu} A^{\mu} \phi^+ \phi^-},\\

\end{array}
\end{equation}
where $Q_t$ and $Q_b$ are the electric charges of the top and bottom
quarks, respectively. Beyond one loop, the following terms
\begin{equation}
{\cal L}_2=\displaystyle{-\frac{g_2 m_t}{2 M_W} H \bar{t} t
           +i\frac{g_2 m_t}{2 M_W} \phi^0 \bar{t}\gamma_5 t}
\end{equation}
should also be added. Here we have ignored the small quark mixing. Note that
only the left-handed component of the $b$-quark couples to the top quark,
so that taking $m_b=0$ is safe and will not produce collinear
or mass singularity because the collinear configuration is forbidden by the
conservation of angular momentum.

Now we analyze the Lorentz structure of the one-loop amplitude for
the process $\gamma(k_1,\epsilon_{\mu}^{(1)})\gamma(k_2,\epsilon_{\nu}^{(2)})
\rightarrow b(p_1) \bar{b}(p_2)$ from the following physical requirements:
$(1)$ on-shell conditions, $k_i^2=p_i^2=0,~\rlap/p_2 v=0=\bar{u}\rlap/p_1$;
$(2)$ terms proportional to $k_{1\mu}$ or $k_{2\nu}$ being automatically
cancelled and thus dropped from the beginning;
$(3)$ left-handedness of the $b$. It is then straightforward to
write down the complete set of independent structures for
the amplitude,
\begin{equation}
\begin{array}{l}
i{\cal A}_{\mu\nu}^{1-{\rm loop}}
=\displaystyle{\frac{ie^2}{(4\pi)^2} \frac{G_F m_t^2}{2\sqrt{2}}
{\bar u}_L [Q_t^2 A_{\mu\nu}^{(t)}
          +Q_t Q_b A_{\mu\nu}^{(tb)}
          +Q_b^2 A_{\mu\nu}^{(b)}] v_L, }      \\ \\

\displaystyle{A_{\mu\nu}^{(i)}=(\rlap/k_1-\rlap/k_2)
          [g_{\mu\nu} h_1^{(i)}+p_{1\mu}p_{1\nu} h_2^{(i)}
          +k_{2\mu}k_{1\nu} h_3^{(i)}+k_{1\nu}p_{1\mu} h_4^{(i)}
          +k_{2\mu}p_{1\nu} h_5^{(i)} ]  }       \\ \\

{}~~~~~~~\displaystyle{+\gamma_{\mu}(k_{1\nu} h_6^{(i)}+p_{1\nu} h_7^{(i)})
          +\gamma_{\nu}(k_{2\mu} h_8^{(i)}+p_{1\mu} h_9^{(i)})}  \\   \\

{}~~~~~~~\displaystyle{+i\epsilon_{\rho\alpha\mu\nu}\gamma^{\rho}\gamma_5
            (k_1-k_2)^{\alpha} h_{10}^{(i)}},
\end{array}
\end{equation}
where the form factors $h_a^{(i)}$ are functions of the Mandelstam
variables $s,~t,~u$ and are related to each other by crossing symmetry.
{}From the naive dimension counting and the fact that the leading terms
are independent of $\xi_W M_W^2$ and that ${\cal A}^{1-{\rm loop}}$ should
be finite as the energy $\sqrt{s} \to 0$, it is tempting to conclude that
$h_1^{(i)}$ and $h_{6-10}^{(i)}$ would behave as $m_t^{-2}$ in the heavy top
limit, and would thus contribute a leading term of $O(G_F m_t^0)$ in
${\cal A}^{1-{\rm loop}}$. However, {\it this naively expected behaviour
actually does not appear due to an additional constraint from
$U(1)_{\rm e.m.}$ gauge invariance}. To put it simply, gauge invariance
dictates the lowest dimension that a gauge invariant structure should carry
so that the above analysis breaks down. In the present case, the amplitude can
be expanded in a complete set of gauge invariant structures,
\begin{equation}
\begin{array}{l}
i{\cal A}_{\mu\nu}^{1-{\rm loop}}
=\displaystyle{\frac{ie^2}{(4\pi)^2} \frac{G_F m_t^2}{2\sqrt{2}}
 \Sigma_{a=1}^{5}[Q_t^2 f_a^{(t)}+Q_t Q_b f_a^{(tb)}+Q_b^2 f_a^{(b)}]
 {\bar u}_L O_{\mu\nu}^a v_L},           \\  \\

\displaystyle{O_{\mu\nu}^1
  =(\rlap/k_1-\rlap/k_2)(k_{2\mu}k_{1\nu}-g_{\mu\nu}k_1\cdot k_2)},\\ \\

\displaystyle{O_{\mu\nu}^2
  =(\rlap/k_1-\rlap/k_2)(p_{1\mu}p_{1\nu}k_1\cdot k_2
                        -k_{2\mu}p_{1\nu}k_1\cdot p_1
                        -k_{1\nu}p_{1\mu}k_2\cdot p_1
                        +g_{\mu\nu}k_1\cdot p_1 k_2\cdot p_1)}, \\ \\

\displaystyle{O_{\mu\nu}^3
  =2\gamma_{\nu}(-p_{1\mu}k_1\cdot k_2+k_{2\mu}k_1\cdot p_1)
   +(\rlap/k_1-\rlap/k_2)(g_{\mu\nu}k_1\cdot p_1-p_{1\mu} k_{1\nu})},\\ \\

\displaystyle{O_{\mu\nu}^4
  =2\gamma_{\mu}(-p_{1\nu}k_1\cdot k_2+k_{1\nu}k_2\cdot p_1)
   -(\rlap/k_1-\rlap/k_2)(g_{\mu\nu}k_2\cdot p_1-p_{1\nu} k_{2\mu})},\\ \\

\displaystyle{O_{\mu\nu}^5
  =i\epsilon_{\rho\alpha\mu\nu}\gamma^{\rho}\gamma_5
   (k_1-k_2)^{\alpha} k_1\cdot k_2
+(\rlap/k_1-\rlap/k_2)(p_{1\nu} k_{2\mu}-p_{1\mu} k_{1\nu})}   \\ \\

\displaystyle~~~~~~~~~{+k_{2\mu}\gamma_{\nu} (2k_2\cdot p_1-k_1\cdot k_2)
+k_{1\nu}\gamma_{\mu} (2k_1\cdot p_1-k_1\cdot k_2)}.        \\   \\

\end{array}
\end{equation}
Note that $O_{\mu\nu}^{3,4,5}$ are gauge invariant only in the on-shell
sense. $O_{\mu\nu}^{1,2}$ are crossing-odd, $O_{\mu\nu}^{5}$ is
crossing-even and $O_{\mu\nu}^{3,4}$ are crossing-exchanged, so are
their form factors $f_{a}^{(i)}$. One may use alternative sets of structures,
but a nice feature of the above one is that each structure is uniquely
characterized by its first term. Again, by dimension counting and the
finiteness of ${\cal A}^{1-{\rm loop}}$ as $\sqrt{s}\to 0$, we deduce
that, in the heavy top limit, $f_2 \sim m_t^{-6}$,
$f_{a\neq 2}\sim m_t^{-4}$, up to logarithms of the form
$(1+{\rm const.}\ln \frac{\xi_W M_W^2}{m_t^2})$ which take into
account the infrared singularity of box diagrams in the Landau gauge
$\xi_W=0$. Indeed, {\it there are no leading terms, and
${\cal A}^{1-{\rm loop}}$ is then dominated by the next-to-leading
terms of} $O(G_F m_t^{-2} (1+{\rm const.}\ln \frac{\xi_W M_W^2}{m_t^2}))$.

At first sight it seems that the top quark
decouples from $\gamma\gamma\to b\bar{b}$ in its large mass limit.
This is certainly not the case. The heavy top effects are only screened
with leading terms cancelled in observables. To see this we go to higher
loops. The above analysis in terms of form factors applies to the $L$-loop
case after only a slight modification of the factor
$\frac{G_Fm_t^2}{16\pi^2}$ in (3) and (4),
i.e. $\frac{G_F m_t^2}{16\pi^2}\to (\frac{G_F m_t^2}{16\pi^2})^L$.
So for the $L$-loop correction,
\begin{equation}
{\cal A}^{\rm L-loop} \sim O(G_F^L m_t^{2(L-2)} ), ~~~~
{\rm up~to~logarithms.}
\end{equation}
This is totally different from the decoupling of heavy fermions in QED but
is quite similar to the {\it screening} phenomenon in the Higgs sector.

Two comments are in order.

$(i)$ As pointed out above, the next-to-leading term is generally $\xi_{W,Z}-$
dependent. This gives us a lesson that whenever the naively expected leading
term is absent in some observables, we should be careful in simplifying the
computation by ignoring the internal weak-gauge-boson contributions.
Especially, when there are infrared singularities associated with unphysical
Goldstone bosons in the Landau gauge, we must include the contributions from
internal $W,~Z$ bosons to obtain a physical result even just to keep the
first non-vanishing term in the heavy top limit.

$(ii)$ Consider the phenomenology at the photon colliders. Since the
contributions from a virtual top quark are generally suppressed
( or screened ) in $\gamma\gamma$ processes not containing external tops,
heavy top effects induced from physics beyond the SM should also be small.
We have computed the one-loop radiative corrections to
$\gamma\gamma\to b\bar{b}$ from the exchange of charged Higgs $H^{\pm}$
in the two Higgs doublet model. For $m_b=4.5 ~ $GeV ( for tree contribution
only ), $m_t=176~$GeV, $M_{H^{\pm}}=400~$GeV, $\cot\beta=5,~\sqrt{s}=100-400$
GeV, and using the spectrum function of back-scattered laser
light$\cite{telnov}$, we find that the relative shift in the total cross
section is less than $10^{-4}$.

The above analysis applies to other processes as well. For example,
since $m_t$ is the largest scale in the decays $b\to s\gamma$$\cite{inami}$
and $H\to\gamma\gamma$ ( or $gg\to H$$\cite{georgi}$ )
and the one-loop momentum integrals
are seemingly linearly divergent, one would naively expect that the decay
amplitudes
behave as $m_t^2$. Actually this leading behaviour is screened by the
appearance of photons in final states. Due to the $U(1)_{\rm e.m.}$
gauge invariance, the effective Lagrangians are, respectively,
\begin{equation}
\begin{array}{l}
{\cal L}_{\rm eff}^{1}=\displaystyle{A e\frac{m_b}{v}\frac{m_t}{v}
      {\bar s}_L\sigma_{\mu\nu}b_R F^{\mu\nu}}, \\ \\

{\cal L}_{\rm eff}^{2}=\displaystyle{B e^2\frac{m_t}{v}H F^{\mu\nu}F_{\mu\nu}}
, \\
\end{array}
\end{equation}
where a factor of $m_b$ has to appear in ${\cal L}_{\rm eff}^1$
to flip the helicity since we have set $m_s=0$.
$A=a/m_t,~B=b/m_t$, and $a,~b$ are finite pure numbers in the heavy top
limit. Thus this only leads to a next-to-leading behaviour which is constant
in $m_t$. The $m_t^2$ dependence first appears at two-loops$\cite{liao}$, as
argued above.

To summarize, we emphasize the importance of local gauge invariance
in causing the screening of the heavy fermion effects in our discussion.
In spontaneously broken gauge theories like the SM, although the heavy top
quark does not decouple as in QED, its effects may be {\it screened} in low
energy processes involving photons. Intuitively, for processes containing
external photons ( or gluons ), local gauge invariance makes the
photons ( gluons ) carry higher powers of momenta than naively expected, so
that the powers of the heavy fermion mass ( as the heaviest mass scale ) will
be lowered as compared with the naive expectation. This kind of screening is
different from  Veltman's in the sense that the latter is due to
the algebraic symmetry structure in the Higgs sector of the SM$\cite{einhorn}$.

We thank Hong-Yi Zhou for an independent check of the numerical
result and Qing Wang for discussions. This work was supported
in part by the National Natural Science Foundation of China
and the Fundamental Research Foundation of Tsinghua University.


\newpage
\baselineskip=20pt
\begin{thebibliography}{20}
\begin{enumerate}
\bibitem{ac}T. Appelquist and I. Carazzone, Phys. Rev. D {\bf 11}, 2856(1975)

\bibitem{dhf} E. D'Hoker and E. Farhi, Nucl. Phys. B {\bf 248}, 77(1984)

\bibitem{veltman}M. Veltman, Nucl. Phys. B {\bf 123}, 89 (1977)

\bibitem{sikivie}P. Sikivie {\it et al.}, Nucl. Phys. B {\bf 173}, 189(1980)

\bibitem{screening}M. Veltman, Acta. Phys. Pol. B {\bf 8}, 475(1977)

\bibitem{bij}J. van der Bij and M. Veltman, Nucl. Phys. B {\bf 231}, 205(1984);
J. J. van der Bij, {\it ibid}, B {\bf 248}, 141(1984)

\bibitem{einhorn}M. B. Einhorn and J. Wudka, Phys. Rev. D {\bf 39}, 2758(1989)

\bibitem{brodsky}For a review, see e.g., S. J. Brodsky and P. M. Zerwas,
at Workshop on Gamma-Gamma Collider, Berkeley, CA, March 28-31, 1994
( Preprint SLAC-Pub-6571 )

\bibitem{telnov}V. Telnov, Nucl. Instrum. Methods A {\bf 294}, 72(1990)

\bibitem{inami}T. Inami and C. Lim, Prog. Theor. Phys. {\bf 65}, 297(1981)

\bibitem{georgi}H. M. Georgi, S. L. Glashow, M. E. Machacek
and D. V. Nanopoulos, Phys. Rev. Lett. {\bf 40}, 692(1978)

\bibitem{liao}Yi Liao, PhD Thesis, Jan. 1994,
Institute of Theoretical Physics, Beijing ( unpublished );

A. Djouadi and P. Gambino, Phys. Rev. Lett. {\bf 73}, 2528(1994)
\end{enumerate}
\end{thebibliography}
\end{document}